\documentclass{article}
\usepackage[dvips]{graphicx}
\begin{document}
\begin{center}
{\bf Effects of introduction of  new resources  and fragmentation
of existing resources on
limiting wealth distribution in asset exchange models.}\\

M. Ali Saif\footnote{e-mail contact :ali@cms.unipune.ernet.in} 
\\{\em Department of Physics, 
University of Pune, Ganeshkhind, Pune, 411 007,
India}\\ 
 Prashant M.
Gade\footnote{e-mail contact : gade@unipune.ernet.in} 
\\{\em Center for
Modeling and Simulation, University of Pune, Ganeshkhind, Pune, 411 007,
India}\\ 
\end{center}

\begin{abstract}
Pareto law, which states that wealth distribution in societies have a 
power-law tail, has been a subject of intensive investigations in statistical 
physics community. Several  models 
have been employed to explain this behavior.  However, most of the 
agent based models assume the  conservation of number of agents and wealth.
Both these assumptions are unrealistic. In this paper, we  
study the limiting wealth distribution when one or
both of these assumptions are not valid. 
Given the universality of the law, we have tried to study
the wealth distribution from the asset exchange models point of view.
We consider models in which a) new agents enter the market at constant
rate b) richer agents fragment with higher probability introducing 
newer agents in the system c) both fragmentation and entry of new agents
is taking place. While models a) and c) do not conserve total wealth
or number of agents, model b) conserves total wealth. All these models
lead to a power-law tail in the wealth distribution pointing to
the possibility that more generalized asset exchange models could
help us to explain emergence of power-law tail in wealth distribution.
\end{abstract}

%\miketitle

\section{Introduction}
 A century ago, an Italian social economist Pareto collected and studied 
data of 
distribution of income across several European countries. He observed 
that 80\% of the income is in 20\% hands and the distribution of income 
has a power-law tail, {\it{i.e.}} $p(x)\propto x^{-1-\nu}$, where 
$p(x)$ probability that an individual has income $x$ . The exponent $\nu$ 
is called Pareto index. 
The exponent measured by him for different kingdoms and countries varied
between 1.1 to 1.7.  The distribution of wealth also shows a similar behavior.
The validity of Pareto law was questioned and reexamined many times. 
In modern times, the 
Japanese, Australian and Italian personal income distribution 
have been shown to demonstrate a log normal distribution for lower 
income coupled with power-law tail \cite{fuji,aoya,clem}. 
For wealth distribution, the distribution of wealth in rich indian families 
has a power-law tail \cite{sinha}.  Same feature is observed in the wealths of 
Hungarian aristocratic families \cite{hegyi}. 
Even for ancient Egyptian society, it has  been conjectured that 
the wealth distribution had power-law tail \cite{abul}. 
The empirical studies on  data of the distribution of income and 
wealth in modern USA and UK  show a power-law tail  as well\cite{drag}.    
All these studies suggest the existence of
power-law tail of wealth distribution and income distribution in
different societies in different parts of the world and it seems to be
true in older as well as modern times. However,  the value of the 
exponent changes in different societies. 
We will be attempt to explain  wealth distribution
from the viewpoint of asset exchange models in this paper.

Given that the universality of Pareto law is so robust, cutting across 
economies which follow different financial system, and even across time, 
there must be a simple explanation for this feature. Several models
have been proposed from this viewpoint.  There are  attempts 
to use ideal-gas like models which recover these features 
\cite{drag1}. Models in 
analogy with directed polymers in random media \cite{mars}
have been proposed.  Generalized Lotka-Volterra dynamics \cite{solomon}
and stochastic evolution equation which incorporate trading as
well as random changes in prices of investments \cite{bouchaud}
have also been proposed.  

Trading is an economic activity which is common to all 
systems in all countries and has been so from
time immemorial. Thus asset exchange models  
which are simplest models of economic transaction should give us 
an explanation of Pareto law.  There have been several attempts 
in this direction \cite{ispo,chak,chat,ali,chat1}. 
We try to study it from the viewpoint of asset exchange models.
In literature, two types of asset exchange 
models have been studied extensively. In these models, there is neither
consumption nor production of wealth.
One of models is called yard-sale model (YS) and 
other is known as theft-fraud (TF) Model \cite{haye}.
In YS model, the amount at stake is certain
fraction of wealth of poorer agent
while in TF model, both agents put a certain fraction
of their wealth to stake.  However, 
none of these models reproduces the power-law distribution of wealth. The 
YS model leads to condensation of wealth in the hand of one agent 
asymptotically while the TF model which is  an 
ideal-gas like model gives us an
exponential distribution of wealth. In this context,
several approaches have been attempted to reproduce
power-law tail in wealth distribution starting from asset exchange 
models. 
Sinha showed that modified YS rule in which poorer player
wins with higher probability leads to a power-law 
distribution of wealth \cite{sinh1}.     
In a previous paper, we showed that mixing the above two
models leads to power-law tail in resultant wealth distribution \cite{ali}.
Several other variants like introducing
altruism in YS and TF models \cite{achach},
introducing saving propensity in TF type 
models \cite{chak,chat} etc. have been studied.

A common  and rather unrealistic feature of these models is that the rules
do not allow total wealth in a society to fluctuate, nor the 
number of players in the society change in time. It is clear that
change in working population makes an impact on the wealth
distribution of the country. It is also clear that the 
true GDP (Gross Domestic Product) of the world has increased
over time. Thus the total wealth is not conserved. 
In fact, projection of real GDP growth is obtained 
by summing the estimates of the
percentage changes in: increased labor inputs,
increased capital inputs and
productivity growth \cite{Baumol}. 
Thus it is clear that increase in labor supply will increase
the real GDP (the proportionality constant is elasticity 
of output with respect to labor).  In most societies, 
this number keeps growing. (Of course, there are exceptions to this
rule of thumb. In older times, catastrophic events like drought,
earthquake, plague  have wiped out populations. 
In modern economies, even decrease in labor supply
is possible. The populations in countries like Russia, Italy, Ukraine etc.
are reducing and economists are discussing its consequences \cite{alesina}. 
However, this is a new trend and its outcome will be apparent only
after a decade or so \cite{note}.)
Labor supply depends on demography.  Kitov in
has postulated that the
GDP growth rate depends on relative change in the number of people with
a specific age (9 years in the USA) \cite{www2}. 
Thus the linear relationship between growth of real GDP and growth of
(working) population is a reasonable assumption.
We are not taking into account the impact of decreasing labor
participation ratio on economy in this paper which is a
new phenomenon.
We are  not taking into account the impact of
productivity or fitness of the agents or influx of new capital 
by agents already in the system.
We will deal with the only effect that 
the total wealth also keeps increasing since
increasing population discovers newer sources of income.
We will systematically analyze the impact of 
increase in number of agents in the Yardsale model of asset exchange.

The wealth often splits at the higher end of the spectrum. Rich people have 
descendents who become independent agents in their own right. Similarly, large 
corporations split into entities which function independently.
There is  a fragmentation of wealth when people have children or 
companies split. Apart from the fact that large wealth is difficult to
manage, there are social and legal pressures which encourage 
division of wealth of rich people. The society at large, resents 
concentration of wealth in hands of some people leading to
income inequalities. The government, on the other hand,
wants to discourage monopolies from the perspective of encouraging
economic efficiency and puts in measures like antitrust act,
ceiling act, quotas 
\cite{antitrust,india}
etc.
These measures affect rich people more
than the poor. 

There are previous attempts to take into account these factors. 
Slanina has given a model with nonconserved wealth
but conserved number of agents.
He models the wealth distribution in analogy with 
inelastically scattering particles and  reports a 
power-law distribution with Pareto
index in the interval [1,2] depending on a free parameter
introduced in the model \cite{slanina}.
There is an attempt to take into account splitting of wealth
between agents. 
R. Coelho et al introduced the  family-network model 
for wealth distribution in societies.  Here, they assume
fragmentation of wealth of older agents among 
their neighbors. This agent reappears with zero age and gets linked to 
two randomly selected agents that have wealth greater 
than a minimal value $q$. The wealth $q$ is taken away from the wealth of 
that selected agents and it is redistributed in a random and preferential 
manner  in society. This model leads to Pareto-like power-law tail for 
the upper $5\%$ of the society. The Pareto index in this model found to 
be $1.8$ \cite{coel}. But this model is static, and total 
wealth and  number of agents are conserved.
Lee and Kim introduced the model with nonconserved number of 
agents similar to the model of growing network . In that model, the 
number of agents increases linearly with the time but the model does not 
consider any exchange of assets, 
{\it {i. e.}}  flow of money between the agents. The wealth 
production for any agent is due to intrinsic ability to produce wealth.
This model leads to power-law tail of the wealth distribution \cite{lee}.
As we argued, given the universality of the law, we feel that one should be 
able to obtain it within the paradigm of asset exchange models.
Despite its faults, we believe that YS model is a good model
of financial transactions. 
We make an attempt to explore the `design space' of asset exchange models,
in particular that of YS model, by introducing 
changes in capital and labor.
Taking YS model as a basic model, we investigate
the asymptotic distribution of wealth
with nonconserved of number of agents and/or the total wealth.

We introduce three different models in 
this context. We investigate models mimicking introduction of newer wealth 
(with newer agents),
fragmentation of wealth with newer agents coming into play and a model
in which both the processes are occurring.
 
\section{The Model(s) and Simulations} 
We focus on YS model of asset exchange. In YS model, which we 
believe to be basically correct description of the asset exchange, the rule 
is the following: the wealth exchanged between two players is  fraction 
of the wealth of the poorer player.  Mathematically, we define it as: if we 
have agents $i$ and $j$ have wealth $x_i(t)$ and $x_j(t)$ at time $t$ and 
are chosen to be updated. Their wealth at next time step will be:
 \begin{eqnarray}
 x_i(t+1)=x_i(t)+\Delta x\\
 x_j(t+1)=x_j(t)-\Delta x
 \end{eqnarray}
 
 $\Delta x=\alpha$ ${\rm{min}}(x_i(t),x_j(t))$
 Where $\alpha$ random number in the interval [0,1]. 
The wealth of all other agents does not change $x_k(t+1)=x_k(t)$ for all
$k$ different from $i$ and $j$.
 
As mentioned before,
this model leads to the unrealistic outcome that the entire wealth
is owned by one  agent asymptotically.  As mentioned in the previous 
section, various modifications of this model do not 
yield a satisfactory power-law tail either. We would like to make a change 
which be believe is realistic and has not been taken into account before. 

There are two processes which need to be accounted for: 

a) The economic system
is not closed and newer agents come in bringing  in their own money. As we
stated before it is reasonable to assume that the true GDP growth will
be proportional to labor supply or number of agents.

b) The wealth often splits at the higher end of the spectrum. Rich people have 
progeny who become independent agents in their own right. Similarly, large 
organizations often split into daughter organizations for smoother
management. We explore variants of YS model 
 in which richer agents keep splitting
with higher probability. 
In these models the number of agents is not conserved but increase 
in time (which is very realistic) and see an impact of such a system in the 
wealth distribution pattern. 
We believe YS process where the asymptotic state is a condensate, 
naturally leads to merging.
We explicitly introduce fragmentation proportional to 
the wealth of an agent.
We also  probe  third situation in which
both the events of newer agents entering the market and division
of wealth for some agents  keep happening.

We will study both the effects individually and also study the 
wealth distribution when both of the above effects are introduced.

We start with a pool 
of $N_0=100$ players that have 
wealth selected randomly from uniform distribution 
in the interval [0,1] in all models below. 
We fix the value of $\alpha=0.5$ in 
all cases. (We checked 
that changing value of $\alpha$ to another constant does not
change results. Changing $\alpha$ 
randomly in time,  or assigning a quenched random value of
$\alpha$ to each agent 
does not change the steady state, if any.)  
Probability density of wealth is approximated from a 
histogram with very fine bins.
We have used 
$10^5$ bins of uniform size in all models below except model B where we 
used $10^4$ bins. After the simulation, we compute the
relative wealth of each agent, normalized by total wealth in the 
system. (Due to scale invariance of power-laws, this linear
transformation does not change the nature or exponent of
the power-law tail.) We compute probability histogram of the relative wealth
and normalize it 
by total number of observations as well as by the bin-size to
obtain an estimator for probability density function of
the wealth \cite{Papoulis}. This procedure is followed in all the case below.
This distribution is noted as $P(\bar x)$.

(A) Inflow of newer agents: In this model, we have a steady flow of
newer agents entering the fray with some wealth.
Thus, neither the number of agents nor the total wealth is conserved.
Let  us denote the number of agents after $k$ rounds
of transactions by $N_k$. After each round of transactions,
the number increases by one and thus $N_k=N_0+k$. 
Each round of transactions consists of as many transactions
as the number of agents, giving each agent a chance to have a
couple of transactions on an average. 
After every round, a new agent enters the system,
that agent has wealth selected randomly 
 from uniform distribution in the
interval [0,1]. Transactions take place according to YS 
rule by choosing two agents randomly to make trade.
We will demonstrate that this system indeed reaches 
a steady state with stable asymptotic
probability distribution of wealth.

We start the simulation with  $N_0=100$. 
We find the probability distribution of wealth after $T$
time-steps and observe that it indeed converges asymptotically.
We average over $100$ initial configurations.
Total wealth increases linearly with the number of
agents. However, we find that the wealth of the richest agent
occupies a fraction of total wealth which remains constant
asymptotically. Thus condensation is clearly not reached. In
order to compare the distributions at different times, we
obtain the probability distribution of $\bar x$ where
$\bar x$ is fraction of wealth that a given agent has acquired.
(Since the power-laws are scale invariant, this transformation
does not change the power-law nature of tail.)
 We show the probability 
distribution $p(\bar x)$ after $T=10^4$ and $T=10^5$ time-steps in Fig. 1.
The distribution has clearly converged. We observe that this 
wealth distribution  has a power-law tail 
with exponent $1.5(5)$. 
Pareto index in this model $\nu=0.5(5)$. This curve has lot of fluctuations.
To average them out we define a new variable 
$f(\bar x)=\int_x^{\infty} p(\bar x) d\bar x$  which is proportional to probability of
having wealth greater than $\bar x$. Clearly $f(\bar x)\propto (\bar x)^{-\nu}$ and 
displays
a power-law tail if $p(\bar x)\propto (\bar x)^{-\nu-1}$ for larger $\bar x$.
In the inset,  we plotted $f(\bar x)$ 
as function of $\bar x$ for $T=10^4$ and $T=10^5$ time steps as in 
original figure. 
The figure in the inset demonstrates  
that the probability distribution has indeed converged
to a steady state. 

 \begin{figure}[htp]
 \includegraphics[width=70mm,height=60mm]{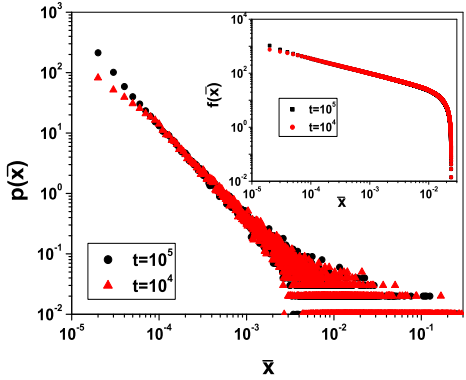}
 \caption{\label{fig1:}We plot wealth distribution function $p(\bar x)$ as
a function of $\bar x$ for model A for $T=10^4$ and $T=10^5$ timesteps.
We clearly see that the model has a steady state which has
a power-law tail with exponent 1.5(5).
 Inset: We plot $f(\bar x)=\int_{\bar x}^{\infty} p(\bar x)d\bar x$ for the
same parameters as the original figure. This clearly demonstrates
that the system has reached a steady state.
}
\end{figure}

(B)Fragmentation of wealth: We consider a situation in which there is 
no inflow of wealth and the richer agents fragment. 
This is a model of
closed society where no extra wealth comes in, but population
increases.  Now inheritances can play a role in which unused
material possessions and assets are divided among siblings. 
However, for poorer agents   
are unlikely to have unused assets and material possessions
which need savings that are translated into investment. 
It is known that savings rate for the rich are higher than
the savings rate for the poor. One more reason why the wealth
is likely to be fragmented at higher end is managerial and
economic efficiency. We also note that 
there are social and legal pressures to stop some agent from
grabbing all the wealth.
Countries have antitrust law and its equivalent to stop
monopolies since monopolies not only lead to higher income
inequality but they also lead to economic inefficiency.
There are land ceiling laws in several countries which essentially
encourage agents to fragment their assets when they are too rich. 
We incorporate this in our model
by saying that rich people divide their wealth with higher 
probability than the poor.
After every $\tau$ time steps, 
we pick an agent randomly and with a probability that is proportional to his 
wealth, we introduce two agents each with half wealth and remove
this agent from the pool. 
(We have checked that change in the value of $\tau$
does not change the asymptotic distribution of
wealth.) 
In other words, if we choose an agent 
(say $k$ th) randomly, we split his wealth  with probability $x_k(t)$.
We reduce wealth of  $k$'th agent  to half 
and introduce the $(N+1)$th agent with same wealth (we checked that 
unequal partitions do not affect our results.) In this model, the number
of agents keeps increasing while total wealth is conserved. Thus,
the average wealth reduces. The probability of fragmentation of
wealth of agent is related to absolute value of his wealth. However,
richer agents keep getting disproportionately targeted for fragmentation. 
Thus the entire distribution slowly tends to a delta function at 
$x=0$ asymptotically with a power law tail which has decreasing weight
as time grows.
However, we may
look for a quasistationary distribution similar to Slanina model \cite{slanina}
in which there is no conserved average wealth but a quasistationary state
with a power law tail.
The tail always shows a 
power-law behavior with exponent $2.8(0)$. 

We carry out simulation with $N_0=100$ agents for $M$ timesteps
and average over $3000$ initial conditions. 
After every $\tau=10$ timesteps, we attempt a fragmentation of
randomly chosen agent as mentioned above.
 The distribution  clearly displays a 
clear power-law tail at any time. For consistency, we have 
displayed distribution of normalized wealth $\bar x$ though 
the total wealth remains constant. (The probability distribution
of $x$ will not be any different except a scale factor.)  We show the wealth 
distribution after $ 10^4, 10^5$ and $10^6$ time-steps in Fig. 2. 
It is clear that the 
distribution has a power-law tail with the exponent $2.8(0)$. 
{\em{This exponent does not change in time.}}  We will support 
this conclusion by a scaling argument.

 %figure1
 \begin{figure}[hpt]
 \includegraphics[width=70mm,height=60mm]{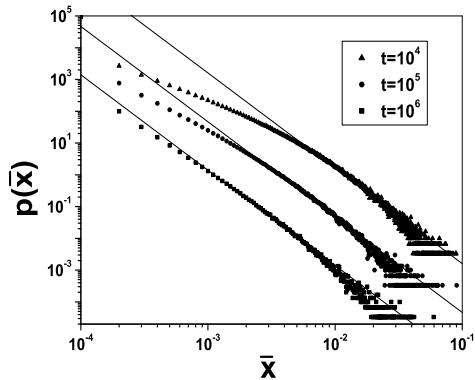}
 \caption{\label{fig2:}We plotted wealth distribution $p(\bar x)$ 
as function of $\bar x$
 for model B for $10^4, 10^5$ and $10^6$ timesteps. We get power-law tail with 
 exponent 2.8(0).}
 \end{figure}

 \begin{figure}[hpt]
 \includegraphics[width=70mm,height=60mm]{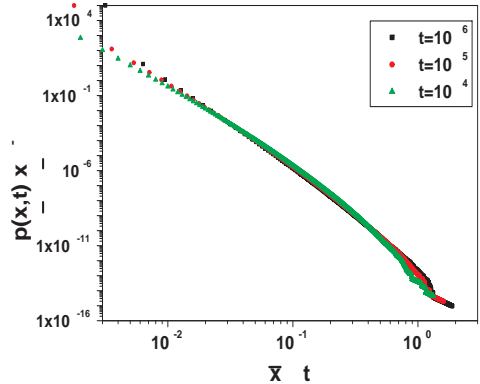}
 \caption{\label{fig3:}The scaling plots of the wealth distribution obtained
 from numerical simulation of model B. The data for three different times
collapse onto a single curve.}
 \end{figure}

The presence of quasistationary state can be demonstrated by
the fact that the distribution admits an interesting scaling behavior.
In Fig. 3, we plot the $p(\bar x,t) \bar x^{-\alpha}$ as a function
of $\bar x^{\alpha} t$ 
for different times. The functional forms at different times
collapse to 
single curve  and  the scaling function is
$p(\bar x,t)\sim \bar x^{\alpha} f(\bar x^{\alpha} t)$ where $\alpha=2+\nu$ and 
$\nu$ the Pareto index in this model. 
 
The exponent $2.8(0)$ 
is certainly more realistic and comparable with exponents obtained  
in empirical studies. The Pareto index is $\nu =1.8(0)$ in this model.
 
(C)Fragmentation and  inflow of agents: This is a model where
both the processes described in model A as well as B happen.
The society is open. Newer agents keep coming in with new
sources of wealth and richer agents keep getting fragmented. Here, 
neither total wealth nor number of agents is conserved.
There are several ways in which this could be achieved.
We have tried three different cases:

 Case I) 
In this case, we couple fragmentation of any agent with addition
of new agent with random wealth chosen from uniform distribution
over [0,1].
We do this to keep 
average wealth constant asymptotically.
Thus every time an agent fragments, we have 
two new agents (one each due to fragmentation and addition)
and on an average wealth of value $1/2$ is introduced in the system
when two new agents are created. (This is due to the fact that
in our models,  any 
agent has initial wealth chosen from uniform random 
distribution
over [0,1]. We do not give any extra wealth to the agent created
due to fragmentation.) 
Thus average wealth which is the first moment of distribution of
wealth will be reaching a constant $1/4$ asymptotically.
We  have checked 
that changing $\alpha$ and making fractions unequal do not change results. 
We start simulation this model with $N_0=100$ agents. After  
carrying out as many YS transactions as the number of agents in the system,
 we randomly select an agent as  
 the model B. We fragment it with probability proportional
to his wealth. Now, we add the new agent with
wealth chosen randomly from interval $[0,1]$ if and only if the 
fragmentation has occurred. We averaged over $100$ initial conditions.
In Fig. 4,  we have shown the probability
distribution of fractional values of wealth $p(\bar x)$
after  $T=10^4$ and   $T=10^5$ timesteps. 
It is clear that  system indeed reaches  a stationary state with power-law tail.
The power-law displayed has an exponent $1.7(7)$.
Pareto index in
this model $\nu=0.7(7)$. As the model A, we plotted $f(\bar x)$
as function of $\bar x$ for $T=10^4$ and $T=10^5$ time steps as in
original figure.
The figure in the inset demonstrates
that the probability distribution has indeed converged
to a steady state.

Case II) Here we add a new agent (with probability one)
after every round of transactions, {\it {i.e.}} 
after as many transactions as the number
of agents in the system. This is very much like Model A.
Additionally, we also choose an agent randomly after every
round of transactions and  fragment it with
probability proportional to his wealth.
In this case, the new agent with his own wealth enters the system at least 
as frequently
 as the event of fragmentation of the older agent (which has
very little probability). Thus we expect a
steady state where the total wealth increases linearly with
number of agents. Again, in order to be able to 
compare distributions of wealth at different times, we obtain
the probability distribution $p(\bar x)$ where $\bar x$ is
normalized value of wealth. The figures clearly demonstrate
a steady state.
We have checked 
also that changing $\alpha$ and making fractions unequal do not change results. 
We start simulation with $ N_0=100$ and average over $100$ 
configurations. In Fig. 5, we demonstrate the probability 
distribution after $T=10^4$ and after $T=10^5$ timesteps.
This model clearly
leads to a distribution with power-law tail with exponent $1.9(0)$. 
 Pareto index in this case $\nu=0.9(0)$.   
As the model A, we plotted $f(\bar x)$
as function of $\bar x$ for $T=10^4$ and $T=10^5$ time steps as in
original figure as an additional evidence for the 
approach to a steady state.

Case III) We have also tried another possibility where 
no particular measures are adopted to make
average asymptotic wealth constant.
It is unrealistic to assume that new agent enters the fray 
only after exactly $N$ transactions are completed. However, it 
is reasonable to assume that number of transactions would be 
proportional to number of agents. Hence, 
we make a probabilistic rule that new agent joins with a probability inversely 
proportional to number of agents in the system at that time. Thus
we allow one agent to join with probability $1/N$ 
after every $\tau$ timesteps where $N$ is number of agents
at that time. We  also  attempt to fragment a randomly agent
chosen with probability proportional to his wealth  after every
$\tau$ timesteps. Though total wealth keeps increasing with
number of new agents joining, it is not necessary that the
increase will be linearly proportional to total number of agents. 
However, our simulations indicate that total number of agents
and total wealth keep increasing linearly with time.
Thus fractional wealth $\bar x$ is a good variable
to analyze.
We start with $N_0=100$ agents.  After every $\tau=10$ timesteps we 
choose an agent randomly and fragment it
with probability proportional to his wealth while a new
agent joins with a probability inversely proportional to 
total number of agents in the system. (We have checked that
changing $\tau$ does not alter the results.)
Here the system does not reach saturation during
our simulation time. However, we observe that the
distribution has a power-law tail.
We have plotted the wealth
distribution $p(\bar x)$
after $T=10^6$ timesteps. It clearly has a power-law tail 
with exponent $2.2(2)$ (See Fig. 6).
After the power-law,
there is a small peak in the distribution at very large masses 
as in case of some particle aggregation models \cite{majumdar}. We average 
over $100$ configuration. Pareto index in this model $\nu=1.2(2)$
at $T=10^6$ time steps.
Though the steady state is not very clear in this situation, the
exponent does not change much at later times. The exponent is high 
and is comparable to realistic societies. 

%figure2
 \begin{figure}[hpt]
 \includegraphics[width=70mm,height=60mm]{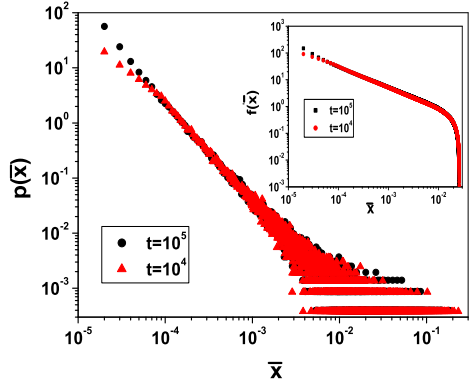}
 \caption{\label{fig4:}We plot wealth distribution $p(\bar x)$ as
a function of $\bar x$ for model C case I for $T=10^4$ and $T=10^5$ timesteps.
We clearly see that the model has a steady state which has
a power-law tail with exponent 1.7(7).
 Inset: We plot $f(\bar x)=\int_{\bar x}^{\infty} p(\bar x)d{\bar x}$ for the
same parameters as the original figure. This clearly demonstrates
that the system has a steady state.}
 \end{figure}

 \begin{figure}[hpt]
 \includegraphics[width=70mm,height=60mm]{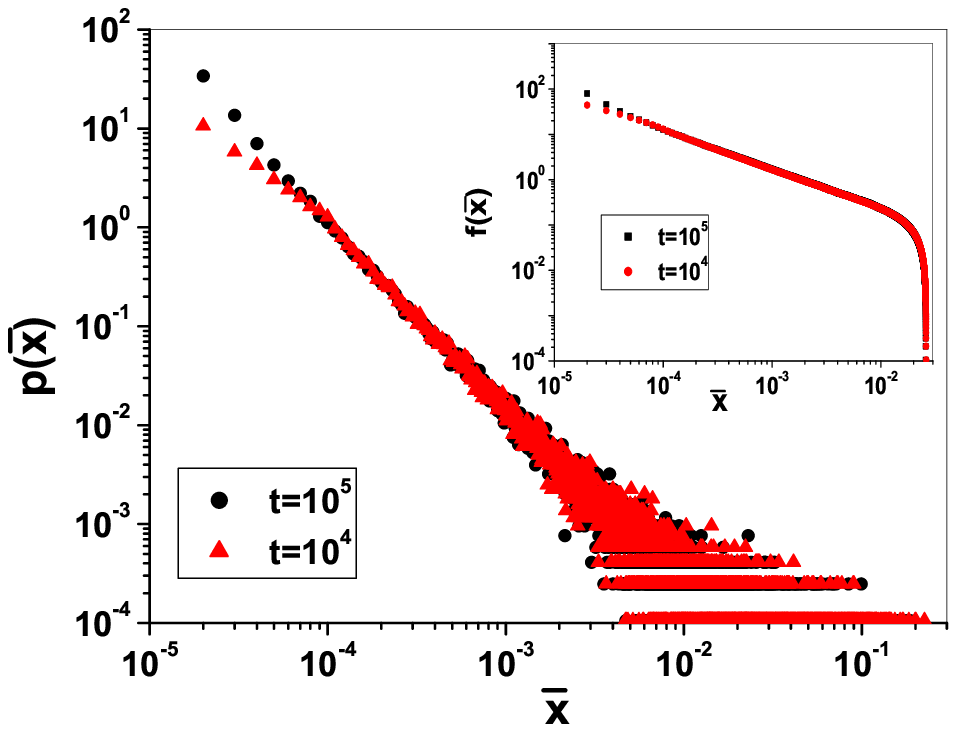}
 \caption{\label{fig5:}We plot wealth distribution $p(\bar x)$ as
a function of $x$ for model C case II for $T=10^4$ and $T=10^5$ timesteps.
We clearly see that the model has a steady state which has
a power-law tail with exponent 1.9(0).
 Inset: We plot $f(\bar x)=\int_{\bar x}^{\infty} p(\bar x)d{\bar x}$ for the
same parameters as the original figure. This clearly demonstrates
that the system has a steady state.}
 \end{figure}

\begin{figure}[hpt]
 \includegraphics[width=70mm,height=60mm]{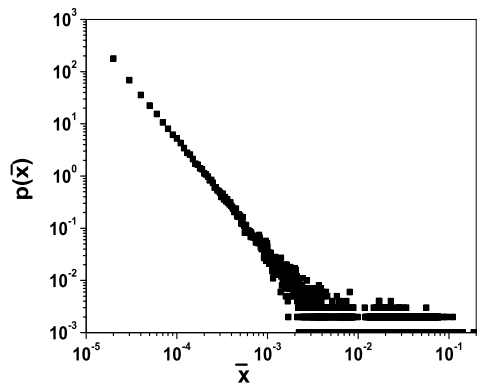}
 \caption{\label{fig6:}Wealth distribution $p(\bar x)$ as function of $\bar x$
 for model C case III for $T=10^6$ timesteps. We get 
power-law  distribution with
 exponent 2.2(2) at this time though there is no clear steady state.}
 \end{figure}

We have checked that the power-law is a better visual
fit than lognormal for
all the cases discussed above.
Besides,  we have checked the goodness of fit by
finding $\chi^2/DoF$ and $R^2$ for three models by fitting it a
power-law functional form and lognormal fit.  The values are given in Table 1.
 It is clear that $R^2$ values are higher and very close to unity for
power-law fit demonstrating  that this model is relevant for higher fraction 
of data. The  $\chi^2/DoF$ values are lower by orders of magnitude
for a power-law fit,  which shows
that error is far smaller for this fit in all the cases.

\begin{tabular}{|l|l|l|l|}
\hline
\multicolumn{4}{|c|}{Table 1: Comparison of Power-law and Lognormal Fits.} \\
\hline
Model & Fit &  $\chi^2/{\rm{Dof}}$ &$R^2$\\ \hline
%\multirow{2}{*}
{Model A} & Lognormal & 5.56 $\times 10^{2}$ & 0.92 \\
 & Power-law &  1.7 $\times 10^{-1}$  & 0.999\\ \hline
%\multirow{2}{*}
{Model B } & Lognormal & 5 $\times10^{-9}$ & 0.04 \\
 & Power-law & 3.5 $\times 10^{-13}$ & 0.999 \\ \hline
%\multirow{2}{*}
{Model C case I)} & Lognormal & 1.0 $\times 10^{5}$ & 3.6 $\times 10^{-6}$\\
& Power-law & 4.8 $\times 10^{-1}$ & 0.999\\
{Model C case II)} & Lognormal & 1.2 $\times 10^{6}$ & 2.8$\times 10^{-7}$\\
& Power-law & 1.87 & 0.999\\
{Model C case III)} & Lognormal & 2 $\times 10^{4}$& $0.17$\\
 & Power-law & 6.0 $\times 10^{-2}$& 0.998\\
\hline
\end{tabular}\\

\section{Results and discussion}

We studied three modifications of the YS model. In all three models, 
we observe power-law tails in  wealth distribution with different exponents. In
the first model, we consider an open system, where one new agent decide to
join to the system with his wealth after each round of transaction. With this
modification of YS, we prevent the condensation of wealth that occurs in the  
pure YS system and find the power-law wealth distribution with Pareto index
$\nu=0.5(5)$. This index is smaller than ones observed
in reality.

In the second model, we prevent the condensation of wealth  
by allowing the richest agents to fragment into two new agents each.
This model was inspired from a similar model trying to
give quantitative explanation of power-law tail in the distribution 
of number of casualties in terrorist attacks observed in empirical
data in several countries.
This model
incorporates fragmentation and merging of terrorist groups.
It assumes that, there are a
specific number of attack units.  (Group of people, weapons, explosives,
machines, or even information, which organizes itself to act is a single
unit.) Each attack unit has an attack strength. At each time step, one
attack unit is chosen with probability which proportional to its
strength.  This  unit chooses to fragment into
smaller groups with some probability $q$  
and coalesces with another attack unit (again chosen with probability 
proportional to its strength) with probability $1-q$.
This model exhibits stationary state with 
power-law distribution for the strength of attack units. The
exponent  is $2.5$ \cite{john}, ({\it {i. e.}} Pareto index $1.5$). 
We believe that merging occurs naturally in YS
model. We introduced the idea of fragmentation proportional to the 
wealth of an agent in 
our model B which gave us 
a power-law wealth distribution with Pareto index 
$\nu=1.8(0)$. This value is comparable to ones observed
in realistic societies.
This model does not have a steady state. But the fact that it admits
an interesting scaling behavior,  shows the presence of a quasistationary
state. 

We have also studied  a third model which incorporates both ingredients
of  bringing in  newer wealth and fragmentation 
of richer agents. These are very realistic features
of societies.  In this model,  we always carry out fragmentation of some
randomly chosen agent with probability proportional to its
wealth. Here we studied three cases. In the first case, 
we coupled the fragmentation of any agent with addition of
new agent having random wealth. This leads to a distribution with
Pareto exponent $0.7(7)$. In the second case, after every round
of transactions, you add an agent and also attempt fragmentation.
This leads to Pareto exponent $0.9(0)$. In the third case, 
we add an agent with probability
inversely proportional to number of agents and also try
fragmenting a randomly chosen agent. 
This leads to distribution with Pareto exponent $1.2(2)$.
Except this case, we have demonstrated that a stationary
or quasistationary state  is reached asymptotically.
We would like to mention that the models are robust with
respect to change in parameters $\tau$ and $\alpha$ and results
do not depend on the precise values of these parameters.

Similar models have been studied in nonequilibrium statistical
physics in the context of aggregation models. Takayasu 
has studied a system in which 
particles are injected at 
a steady rate, they diffuse and try to form an  aggregate. \cite{Takayasu}
Asymptotic state here is known
to have a power-law tail. There are also models in which particles 
chip off from aggregates. (For a survey of these different models,
see \cite{majumdar} ) In our model, the wealth is not discrete,
and YS process is similar to but not the same as  coagulation
of particles. However, some analytic insights could be gained from
the analysis of particle aggregation models and studies are being
carried in this direction. 

MAS thanks Govt. of Yemen for financial 
support and PMG thanks S. Sinha for discussions.

\end{document}